Draft – Comments and Suggestions Welcomed
Version 1.4

Growth in the "New Economy": U.S. Bandwidth Use and Pricing Across the 1990s

Douglas A. Galbi
Senior Economist[1]
Competitive Pricing Division
Common Carrier Bureau, FCC
July 9, 2000

**Summary**

An acceleration in the growth of communications bandwidth in use and a rapid reduction in bandwidth prices have not accompanied the U.S. economy's strong performance in the second half of the 1990s. Overall U.S. bandwidth in use has grown robustly throughout the 1990s, but growth has not significantly accelerated in the second half of 1990s. Average prices for U.S. bandwidth in use have fallen little in nominal terms in the second half of the 1990s. Policy makers and policy analysts should recognize that institutional change, rather than more competitors of established types, appears to be key to dramatic improvements in bandwidth growth and prices. Such a development could provide a significant additional impetus to aggregate growth and productivity.

---

[1] The opinions and conclusions expressed in this paper are those of the author. They do not necessarily reflect the views of the Federal Communications Commission, its Commissioners, or any staff other than the author.

The U.S. economy's performance in the second half of the 1990s was much better than in the first half of the 1990s. Output grew much faster: U.S. real non-farm business output grew at an average annual rate of 4.8% 1996-1999, compared to 2.8% 1991-1995. Productivity growth also showed significant improvement between the first and second halves of the 1990s, with labor productivity growth rising to 2.6% per year 1996-1999 compared to 1.5% per year 1991-1995.[2] Understanding these beneficial developments and their implications for the future is vitally important in formulating a wide range of economic and social policies.

The U.S. economy's performance in the second half of the 1990s has stimulated belief that a "new economy", based on digital information processing and communications, is emerging. Empirical studies show that computer hardware, software, and communications equipment drove U.S. growth in the second half of the 1990s,[3] while the Internet and electronic commerce have attracted much public attention.[4] The preface to a report entitled "Digital Economy 2000" notes: "…confidence has increased among experts and the American public that the new, proliferating forms of e-business and the extraordinary dynamism of the industries that produce information-technology products and services are harbingers of a new economic era."[5] Another report associates over a half trillion dollars of U.S.-based company revenue in 1999 with what it calls the Internet Economy.[6] The European Union has conceptualized impending economic changes in terms of an Information Society, and it has launched an "eEurope" initiative to ensure that the European Union benefits fully from developments it considers "the most significant since the Industrial Revolution".[7]

The U.S. growth experience in the 1990s and the perceptions of a new economy contrast sharply with trends in bandwidth use and pricing across the 1990s. Dramatic increases in bandwidth use and dramatic reductions in bandwidth prices that were predicted to occur in the second half the 1990s did not occur. Over-all bandwidth use has grown robustly throughout the 1990s, but growth has not significantly accelerated in the second half of

---

[2] See Oliner, Stephen D. and Daniel E. Sichel, "The Resurgence of Growth in the Late 1990s: Is Information Technology the Story?" Working Paper (May, 2000) [available on the web at http://www.federalreserve.gov/pubs/feds/2000/200020/200020pap.pdf], Tables 1-2.

[3] Oliner and Sichel (2000) and Jorgensen, Dale W. and Kevin J. Stiroh, "Raising the Speed Limit: U.S. Economic Growth in the Information Age," Working Paper (May, 2000) [available on the web at http://www.economics.harvard.edu/faculty/jorgenson/papers/dj_ks5.pdf].

[4] Prior to 1995, scholars and students were the primary users of the Internet. In May 2000, 82.6 million persons in the U.S. used the Internet. See http://209.249.142.27/nnpm/owa/NRpublicreports.usagemonthly

[5] U.S. Department of Commerce, Economics and Statistics Administration, *Digital Economy 2000* (June 2000), preface [available at http://www.ecommerce.gov].

[6] Cisco Systems and the University of Texas, Measuring the Internet Economy (June 6, 2000) [on the web at http://www.internetindicators.com ].

[7] eEurope: An Information Society For All. See http://europa.eu.int/comm/information_society/eeurope/index_en.htm

For an interesting, fact-based argument that computers and the Internet are not as significant as past great inventions, see Gordon, Robert J., "Does the `New Economy' Measure up to the Great Inventions of the Past?" May 1, 2000 draft of a paper for the *Journal of Economic Perspectives*, available at http://faculty-web.at.northwestern.edu/economics/gordon/351_text.pdf

1990s. Reductions in bandwidth prices, in sharp contrast to reductions in computer prices, have not accelerated in the second half of the 1990s. With respect to communications bandwidth, the development of a new economy, and any associated impact on macroeconomic growth and productivity, appear to be still yet to come.

**I. Bandwidth Expectations, Technology, and Use in the 1990s**

Influential industry observers have long anticipated dramatic changes in communications bandwidth. In an article published in 1994, George Gilder foresaw a "bandwidth tidal wave," a "tsunami of gigabits."[8] Microsoft's Bill Gates declared in October 1994, "We'll have infinite bandwidth in a decade's time." In the same year, Andy Grove, Chairman of Intel, stated, "If you are amazed by the fast drop in the cost of computing power over the last decade, just wait till you see what is happening to the cost of bandwidth." Long-time industry observer Jack Rickard, then editor of a highly regarded trade journal and known for debunking spin and hype, declared in 1996, "…bandwidth across the board will increase by at least one order of magnitude every two years." He predicted that in the year 2000 U.S. backbone bandwidth would be 5 Gbps, while home users would have bandwidth of 2.88 Mbps.[9]

Developments in optical technology unquestionably have made massive increases in bandwidth possible.[10] An October, 1998 Nortel technology newsletter reported that, since 1996, Nortel had shipped more than 1,500 10 Gbps transport systems.[11] In a recent column the president of Lucent's Optical Networking Group stated that over the past decade, optics' price-performance ratio has improved 100-fold.[12] To get a sense of the current state of the technology, consider these facts. An optical transport system scheduled for volume shipments in the third quarter of 2000 uses dense wavelength division multiplexing (DWDM) to provide, without opto-electronic regeneration, a total of 560 Gbps across a single 3600 km optical fiber.[13] Routers with a rich set of packet-forwarding functions are currently available that provide 160 Gbps throughput using a box that is half the size of the typical telecom equipment rack.[14] An all-optical wavelength router, scheduled to be commercially available in December 2000, will handle 256 40 Gbps data channels.[15]

---

[8] Gilder, George, "The Bandwidth Tidal Wave," Forbes ASAP (Dec. 5, 1994) [available on the web at http://www.forbes.com/asap/gilder/telecosm10a.htm]. The subsequent quotes from Gates and Grove are from this article.
[9] Richard, Jack, "Editor's Notes: Bandwidth Arithmetic and Mythology," *Boardwatch* (May 1996), available on the web at http://boardwatch.internet.com/mag/96/may/bwm1.html
[10] See Gilder, George, "Fiber Keeps Its Promise," Forbes ASAP (April 7, 1997). Gilder stated, "The law of the telecosm ordains that the total bandwidth of communications systems will triple every year for the next 25 years."
[11] Nortel Networks, "Pushing the Limits of Real-World Optical Networks," Nortel's Technology Perspectives (October 19, 1998), p. 4 [available on the web at http://www.nortelnetworks.com/corporate/technology/tech_features/collateral/ntp1.pdf ]
[12] This improvement is contrasted with a 60-fold improvement in the price-performance ratio for microprocessors over the same period. Butters, Gerry, "Welcome to Photon Valley" [on the web at http://www.lucent.com/netrev/Viewpoints/ButtersViewpoint.html ].
[13] Nortel Network's (Qtera subsidiary) ULTRA system. See http://www.nortelnetworks.com/corporate/news/newsreleases/2000a/03_07_0000130_ultra.html
[14] Juniper Networks' M160 Internet backbone router. See http://www.juniper.net/products/m160-l2.html

While technological developments have enabled vast increases in bandwidth, the existence of such technology does not necessarily imply its widespread deployment in wide-area communications networks. New transmission technologies work most effectively over new fiber strands that have enhanced optical properties. Thus new fiber deployment indicates an expansion of potential capacity via both an increase in the gross volume of installed fiber and an improvement in its technological vintage. Data indicate that growth in fiber miles deployed in the U.S. has been falling for over a decade. Table 1 shows that fiber miles grew about 25% per year in the beginning of the 1990s, while they grew about 18% per year at the end of the 1990s. The explanation for this trend is clear from Table 1: local incumbents, who currently account for about two-thirds of total fiber miles, have reduced their rate of deployment of new fiber miles.[16] The trend in fiber miles deployed suggests that growth in potential capacity has not accelerated in the second half of the 1990s.

**Table 1**
**U.S. Fiber Miles**
(in thousands)

| Year | Local Incumbents miles | grth. | New Local Entrants Miles | grth. | Long Distance miles | grth. | Total miles | grth. |
|---|---|---|---|---|---|---|---|---|
| 1988 | 1,754 | | | | 1,722 | | 3,476 | |
| 1989 | 2,255 | 29% | | | 1,892 | 10% | 4,147 | 19% |
| 1990 | 3,181 | 41% | 50 | | 2,085 | 10% | 5,315 | 28% |
| 1991 | 4,389 | 38% | 76 | 53% | 2,203 | 6% | 6,668 | 25% |
| 1992 | 5,863 | 34% | 116 | 52% | 2,227 | 1% | 8,206 | 23% |
| 1993 | 7,508 | 28% | 211 | 82% | 2,291 | 3% | 10,010 | 22% |
| 1994 | 9,018 | 20% | 365 | 73% | 2,456 | 7% | 11,839 | 18% |
| 1995 | 10,698 | 19% | 640 | 76% | 2,585 | 5% | 13,924 | 18% |
| 1996 | 12,343 | 15% | 1,306 | 104% | 2,940 | 14% | 16,588 | 19% |
| 1997 | 14,017 | 14% | 1,826 | 40% | 3,419 | 16% | 19,262 | 16% |
| 1998 | 16,077 | 15% | 3,038 | 66% | 3,681 | 8% | 22,796 | 18% |
| 1999 | 17,885 | 11% | 4,739 | 56% | | | | |

Notes and sources: See Appendix B.

---

[15] Lucent Technologies' WaveStar LambdaRouter. See http://www.lucent-optical.com/press/lambdarouter.html . For a useful review of technological developments, see also the Australian Government's *National Bandwidth Inquiry*, Appendix 5 [on the web at http://www.noie.gov.au/projects/information_economy/bandwidth/index.htm ].

[16] Growth in long distance fiber miles, as well as route miles, has increased in the second half of the 1990s. The growing importance of long-term leasing of long-distance fiber may contribute to double-counting in fiber statistics. On the other hand, fiber deployment by non-carriers such as electric utility companies and cable television companies is not included in Table 1. Nonetheless, in the 1990s local incumbents' fiber stock was much larger than that of other entities, hence local incumbents' fiber deployment largely determined the aggregate fiber deployment trend.

More significantly, available data for bandwidth in use do not show a rapid upsurge in the second half of the 1990s. From 1989-1994, regional Bell Operating Company (RBOC) inter-office non-switched bandwidth sold grew about 33% per year, the same growth rate subsequently experienced 1995-1999.[17] See Table 2. Data on combined U.S. trans-Atlantic and trans-Pacific bandwidth, which includes bandwidth that all companies provide, shows growth of 47% per year and 79% per year in 1989-1994 and 1995-1999, respectively. Much of the growth in the latter period, however, depends on the 1999 bandwidth estimate; growth for the period 1995-1998 is 51% per year.[18]

While network providers other than local incumbents have been active in supplying domestic bandwidth, new local providers are not large enough to dramatically affect the growth rates for domestic bandwidth in use. New local network providers earned about 12% of total local leased line revenues in the U.S. in 1998.[19] Surveys of end users and network build-out in major U.S. cities suggest that three or more major new companies are providing leased line services in large cities, and companies other than the incumbent local exchange company (LEC) provided about a third of channel termination facilities on a DS1 equivalent basis in those cities about 1998.[20] However, even if the bandwidth in use from new local network providers grew from zero in 1995 to an amount equal to that of the RBOCs in 1999, the aggregate bandwidth growth rate from 1995 to 1999 would be only about 58% per year.[21]

---

[17] The RBOCs are the largest incumbent local exchange companies (LECs) in the U.S.; they serve about 80% of local switched access lines.

[18] As Rood, Hendrick, "Indicators for bandwidth demand," *Telecommunications Policy* 24 (April 2000) 263-270 points out, U.S. international capacity statistics have significant weaknesses. U.S. international private line revenue statistics are particularly suspect because the revenue associated with international private line connectivity, such as international Internet peering, is not easily accounted for. As Rood notes, the high capacity circuits counts associated with these revenue statistics appear to be inaccurate. Aggregate international capacity figures can be highly distorted by the treatment of intermediate links. For this reason I have focused on aggregate capacity reported by discrete trans-Atlantica and trans-Pacific cables.

[19] *Federal-State Joint Board Monitoring Report*, FCC CC Docket 96-45 (June 1999) Tables 1.5 and 1.6. This figure is based on revenue for local private lines and special access service sold for resale and to end users. The ratio of CLEC to ILEC revenues is divided by the ratio of RBOC to ILEC revenues from 1998 SOCC figures for local private line and special access revenues.

[20] In 1998 and 1999, RBOCs submitted petitions for forbearance from regulations of their leased lines in numerous cities and regions of the U.S. See FCC CC Dockets 98-157, 98-227, 99-1, 99-24, 99-65. These petitions included studies by Quality Strategies documenting the growth of other providers of leased lines. According to the Quality Strategies' survey of end users in Phoenix in the fourth quarter of 1997, companies other than U S West provided 27.9% of channel termination facilities on a DS1 equivalent basis. Quality Strategies, *High Capacity Market Study for Phoenix MSA* (Aug, 1998), submitted with Petition of U S West for Forbearance from Regulation as a Dominant Carrier in Phoenix, CC Docket 98-157. Quality Strategies found a market share for other companies of 47% (channel termination facilities on a DS1 equivalent basis) across a weighted average of five major Texas cities in the second quarter of 1998. See Petition of SBC Companies for Forbearance, CC Docket 98-227 (Dec. 7, 1998), App. A and B. The ensuing record raised significant questions about Quality Strategies' market share data. See *Memorandum Opinion and Order*, FCC 99-365 (rel. Nov. 22, 1999), para. 25-29. In Texas, the non-incumbent LECs' revenue share of intrastate leased lines rose from 12% in 1995 to 49% in 1997, while incumbent LECs' revenue for those services remained roughly constant across that period. See Public Utility Commission of Texas, *Report to the 76th Texas Legislature, Scope of Competition in Telecommunications Markets of Texas* (January, 1999), Tables 12 and 24. These data are based on a Texas PUC survey.

| Year | Table 2 Bandwidth in Use | | | |
|---|---|---|---|---|
| | U.S.Trans-Atlantic & Trans-Pacific | | U.S. incumbent (RBOC) inter-office | |
| | Gbps | Yr-to-Yr Growth | Gbps | Yr-to-Yr Growth |
| 1989 | 2.1 | | 330 | |
| 1990 | 2.1 | 0% | 475 | 44% |
| 1991 | 3.1 | 53% | 683 | 44% |
| 1992 | 6.5 | 108% | 835 | 22% |
| 1993 | 9.4 | 44% | 1,014 | 21% |
| 1994 | 14.3 | 51% | 1,380 | 36% |
| 1995 | 20.6 | 44% | 2,362 | 71% |
| 1996 | 32.2 | 56% | 2,693 | 14% |
| 1997 | 36.0 | 12% | 3,461 | 29% |
| 1998 | 70.9 | 97% | 5,150 | 49% |
| 1999 | 210.2 | 197% | 7,407 | 44% |
| 2000 | 868.3 | 313% | | |

Notes and sources: See Appendix B.

Growth of bandwidth in use for Internet traffic has been dramatic since 1995, but Internet bandwidth is only a small part of total bandwidth in use. A careful study suggests that Internet backbone traffic grew 1000% per year in 1995 and 1996, but the growth rate fell to 100% per year in 1997 and 1998. Total Internet backbone bandwidth in mid-1998 was probably about 110 Gbps.[22] As Table 2 shows, total Internet bandwidth is only about 2.1% of total RBOC interoffice bandwidth in use in 1998.[23] Thus the growth of total bandwidth in use is thus far not greatly affected by the rapid growth of Internet bandwidth.

The evidence and analysis above contrasts not only with expectations of industry leaders in the early 1990s, but also with some current perceptions of a bandwidth explosion.

---

[21] It is worth noting that Quality Strategy studies showed that competitors had significant market share as early as First Quarter, 1994. Quality Strategies estimated that in 1Q1994 Bell Atlantic's competitors had from 25-35% of DS1 equivalent market share in Philadelphia, Pittsburg, Baltimore, and Washington, DC. See Attachment 2, Letter from Raymond Smith, Bell Atlantic CEO to Reed Hundt, FCC Chairman, March 23, 1995.

[22] Odlyzko, Andrew, "The Current State and Likely Evolution of the Internet" [on the web at http://www.research.att.com/~amo/doc/recent.html ].

[23] As Andrew Odlyzko has pointed out in a personal communication, average capacity utilization of Internet backbone links may be about 5 times greater than (local) inter-office links. In addition, the inter-office bandwidth figure is the sum of all link bandwidth, while the Internet bandwidth figure is based on an estimate of 2.5 links traversed per packet. Thus, more comparably measured, the ratio of Internet bandwidth to total inter-office bandwidth might be about 15% in mid-1998.

Citing a statistic that has become more general and more authoritative over time, Reed Hundt, who served as chairman of the U.S. Federal Communications Commission (FCC) from 1993 to 1997, stated in his recent book, *You Say You Want a Revolution*, "In 1999 data traffic was doubling every 90 days, as connected personal computers spread across the globe."[24] Doubling every 90 days implies an annual growth rate of 1,663% per year. An early citation of a similar statistic was the U.S. Department of Commerce's April 1998 report, *The Emerging Digital Economy*, which stated, citing an Inktomi 1997 White Paper, "UUNET, one of the largest Internet backbone providers, estimates that Internet traffic doubles every 100 days."[25] John Sidgmore, Chairman of UUNet and Vice-Chairman of Worldcom, recently stated that UUNet's traffic in the U.S. has been growing 750% per year consistently since 1995.[26]

Such claims, which have expanded in generality and authority with repetition and distance from the source, have provided unwarranted support for the view that a communications bandwidth revolution has helped to create a "new economy" in the U.S. in the second half of the 1990s.[27] Recent evidence indicates that UUNet's traffic growth claims are based on a rather peculiar measure of traffic.[28] Consideration of the more comprehensive and better documented evidence put forward in this section suggests a much different picture of bandwidth growth. The preponderance of evidence indicates that there has not been a dramatic upsurge in total U.S. bandwidth in use in the second half of the 1990s.

**II. Trends in Bandwidth Prices Across the 1990s**

A rapid decline in computer prices is an important fact closely associated with the development of a new economy. The decline in U.S. computer prices accelerated to 28% per year 1995-1998 from 15% per year in 1990-1995.[29] Performance-price trends for computers are popularly described more abstractly in terms of Moore's Law: the performance-price ratio for processors doubles every 18 months.[30] In the second half of the 1990s, the period for doubling the performance-price ratio appears to have shortened to 12 months.[31] These developments indicate high productivity growth in the production of information technology and encourage investment in information technology capital throughout the economy.

---

[24] Hundt, Reed E., *You Say You Want A Revolution* (Yale University Press: 2000) p. 224. No source is provided for the statistic. See also Telecommunications @ the Millennium: The Telecom Act Turns Four, OPP FCC (Feb. 8, 2000) p. 4 [on the web at http://www.fcc.gov/opp/ ].
[25] Chapter 2. Available on the web at http://www.ecommerce.gov
[26] Keynote address at SuperComm 2000, June 5, 2000, as reported in Communications Daily, June 6, 2000.
[27] Odlyzko, op. cit.
[28] According to Rood, op. cit. p. 269, a UUNet source indicated that "…after long internal discussions inside UUNet, it was decided to use the bandwidth-distance product as a measure for the size of the network. This means that when a network of 100 route miles is expanded to a network of thousand route miles, and the new equipment installed on that route is operating at the same bit-rate as the original network, it is considered as a 10-fold increase of the size."
[29] Jorgenson and Stiroh (2000), Table 1. These price statistics are based on an official data series that adjusts for quality changes.
[30] As Coffman and Odlyzko (2000) note, Moore's law as originally put forth by Moore related to how rapidly the number of transitors per microprocessor chip would double. Now Moore's law is generally used to describe rapid improvements in computers' performance-price ratio.

The performance-price ratio for leading-edge optical communications technology also is improving rapidly. Table 3 shows the investment cost per Mbps of capacity on transatlantic optical transport systems going into operation between 1988 and 2000. The investment cost per Mbps of capacity fell by about a factor of 10 from 1989-1995 and from 1995-2000. Note that there does not appear to have been a significant difference in the reduction in investment cost per Mbps from the first half of the 1990s to the second half. Note also that investment cost per Mbps is not the same as the end-user price per Mbps. The latter, and not the former, is much more relevant to bandwidth use.

**Table 3**
**Optical Technology in Trans-Atlantic Cables**

| Year | Cable | Cost ($ mil) | Capacity (Gbps) | Unit Cost ($ mil/Gbps) |
|---|---|---|---|---|
| 1988 | TAT 8 | 360 | 0.5 | 744.0 |
| 1989 | PTAT 1 | 400 | 1.1 | 367.4 |
| 1992 | TAT 9 | 406 | 1.0 | 419.6 |
| 1992 | TAT 10 | 300 | 1.5 | 206.7 |
| 1993 | TAT 11 | 280 | 1.5 | 192.9 |
| 1995 | TAT 12/TAT-13 | 756 | 23.2 | 32.6 |
| 1997 | Gemini | 520 | 23.2 | 22.4 |
| 1998 | Atlantic Crossing (AC-1) | 850 | 61.9 | 13.7 |
| 2000 | TAT-14 | 1500 | 495.5 | 3.0 |
| 2001 | Level 3 | 600 | 247.7 | 2.4 |
| 2001 | Hibernia | 630 | 123.9 | 5.1 |

Source: See Appendix B.

Bandwidth prices do not appear to have fallen significantly for most bandwidth users in the second half of the 1990s. Table 4 shows average nominal prices per Mbps, on a distance-standardized basis, for four types of inter-office circuits that local exchange carriers offer.[32] Prices decreased significantly from 1990-1994 but only slightly from 1995-2000.[33] Circuits that connect customers to telephony company offices ("channel

---

[31] Gordon (2000), p. 10.
[32] These indices are based on all inter-office circuits of the given type that Bell Atlantic (south) and USWest sold in the given year. See Appendix A. The circuit types are VG, a voice grade circuit; DDS 56Kbps, a "digital data service" circuit that provides a low-bandwidth connection designed to provide high-quality data transmission; and DS1 and DS3 circuits, which carry either bundles of voice grade circuits or data traffic.
[33] The FCC has made significant efforts over a long time to promote competition for inter-office circuits. In a series of orders beginning in October, 1992, the FCC required local exchange carriers to offer inter-office transport on a flat-rated, unbundled basis and required large local exchange carriers to offer collocation in their end offices. Moreover, significant collocation has existed for a long time. One legacy of the AT&T divestiture is that AT&T gained points of presence (POPs) collocated in RBOC offices. In

terminations" or "circuit tails") show even less of a price decline. See Appendix Table A2. Given the large differences in bandwidth and price per Mbps among the different types of circuits in Table 4, shifts among circuit types could have a major affect on average prices for bandwidth. But whether constructed using bandwidth weights or revenue weights, an aggregate index covering VG, DDS (56Kbps), DS1, and DS3 circuits shows only a slight decline in the second half of the 1990s.[34] Such a price trend is much different than the sharp and accelerating declines in computer prices across the 1990s.

**Table 4**
**U.S. Local Inter-Office Circuit Prices**
(dollars per month per Mbps)

| Year | VG (64, 128Kbps) | DDS (56Kbps) | DS1 (1.5Mbps) | DS3 (44.7Mbps) | Over-all (bw wtd) | Over-all (rev wtd) |
|---|---|---|---|---|---|---|
| 1990 | 159 | 2,514 | 191 | 16.17 | 161 | 347 |
| 1991 | 156 | 2,421 | 174 | 19.03 | 122 | 361 |
| 1992 | 140 | 2,092 | 177 | 18.07 | 107 | 352 |
| 1993 | 121 | 1,399 | 136 | 16.21 | 79 | 265 |
| 1994 | 134 | 971 | 123 | 17.95 | 69 | 202 |
| 1995 | 140 | 1,171 | 114 | 17.15 | 66 | 276 |
| 1996 | 144 | 900 | 116 | 16.53 | 63 | 219 |
| 1997 | 138 | 846 | 118 | 16.46 | 65 | 233 |
| 1998 | 149 | 925 | 114 | 16.73 | 63 | 231 |
| 1999 | 147 | 942 | 113 | 17.49 | 60 | 219 |
| 2000 | 143 | 878 | 112 | 17.17 | 56 | 185 |

Notes and source: See Appendix B.

While local exchange carriers offer a variety of circuits and services and a large number of rate elements, those discussed above have been the most important on a revenue basis. Throughout the 1990s VG, DDS, DS1, and DS3 circuits have accounted for over 80% of RBOC leased-line revenue. See Table 5. In the beginning of the 1990s the largest component of the "other" category was telegraph service; by the end of the 1990s the largest component of the "other" category was various types of SONET services. Making an interstate leased line operational typically includes a variety of installation

---

1992 28% of AT&T's POPs were collocated in LEC offices. See AT&T Ex Parte Presentation, Dockets 91-213, 91-141, letter from Thomas H. Norris to FCC Secretary, July 17, 1992, attachment. Reporting under the FCC's expanded interconnection rules show considerable additional collocation by early 1997. For example, as of June, 1997 Bell Atlantic (south) had 12 virtually collocating companies and 8 physically collocating companies with a total of 124 virtual collocation arrangements and 35 physical collocation arrangements. These arrangements do not appear to include any AT&T collocated POPs. See Letter from Joseph Mulieri of Bell Atlantic to William Caton, Acting FCC Secretary, re: CC Docket No. 91-141, July 9, 1997.

[34] In constructing price indices such as the U.S. Consumer Price Index (CPI), revenue weights are used to aggregate prices for different products. The 1995 over-all revenue weighted price is strongly influenced by an unusual pricing pattern for Bell-Atlantic DDS 56Kbps in 1995. That pricing may have been an (unsuccessful) attempt to shift customers away from DDS 56Kbps circuits.

charges, cross connect charges, and other charges that a local exchange carrier might establish. However, in revenue terms such charges have not been large; monthly rate elements for channel terminations, inter-office links, and inter-office mileage account for about 85% of RBOC leased line revenue. Thus price trends for these rate elements are a good indication of overall price trends for local exchange carrier leased lines.

| **Table 5** **RBOC Leased Line Revenue in 1989 and 1999** | | |
|---|---|---|
| | Revenue | Shares |
| Circuit Type | 1989 | 1999 |
| VG | 52.9% | 3.5% |
| DDS | 8.1% | 11.6% |
| DS1 | 35.1% | 47.3% |
| DS3 | 1.7% | 19.0% |
| Other | 2.2% | 18.7% |
| | Revenue ($ millions) | |
| Category | 1989 | 1999 |
| Interstate | 1,957 | 5,042 |
| Total | 3,069 | 7,673 |
| Source: See Appendix B. | | |

As Table 5 shows, significant changes in the mix of circuits have taken place across the 1990s. The revenue share of VG circuits has fallen dramatically, while the revenue shares of other circuit types have risen. While Table 4 shows that the price per Mbps is much smaller for larger capacity circuits, developments do not simply reflect a shift toward higher capacity circuits. The revenue shares for low capacity DDS circuits (primarily 56Kbps) has increased sharply across the 1990s. DS1 circuits still have by far the largest share of revenue and currently account for almost half of all leased line revenue. Although the revenue share for digital circuits with capacity greater than 44.7Mbs has risen over time, the revenue share of such circuits is currently only about 18%.[35] These developments suggest that technology that dramatically reduces the price per Mbps for large circuits is not the key to reductions in the price per Mbps that typical users pay.

While the above average prices are for RBOC-provided circuits, new local competitors are not likely to be offering dramatically lower prices for similar circuits. Supplying local wireline bandwidth involves high sunk costs and low marginal costs. Firms with such cost characteristics have a strong incentive to avoid direct price competition. Thus, for example, competitors have largely shunned providing competing switched voice transport between local exchange carrier offices.[36] This service is well established, with little opportunity for competitive differentiation. On the other hand, the absence of a well-recognized, neutral interconnection lattice in the industry means that leased line products are predominately structured as highly idiosyncratic end-point to end-point connections. Given the huge number of possible end-point to end-point combinations, the average level of price competition across all these possible combinations is likely to be quite low.[37] While there has been significant competitive entry in providing point-to-point bandwidth, the overall effect on average bandwidth prices probably has not been dramatic.

---

[35] Note that relatively high capacity circuits have been available for a long time; as early as 1991 USWest was selling 1.244 Gbps channel terminations.
[36] RBOC annual access filings indicate that in 1998 only 3.5% of RBOC-terminated switched voice traffic was carried inter-office on competitors' facilities.

Establishing price trends for long distance bandwidth is much more difficult than establishing price trends for local bandwidth. Because long distance transport rates are not regulated in the U.S.,[38] no comprehensive data are available on them. Prices for long distance leased lines in the U.S. vary widely and are often negotiated on a case-by-case basis with significant discounts available to selected customers.[39] Attempts to create bandwidth exchanges reflect the fact that the current market for bandwidth has high transaction costs and prices that only weakly signal broad industry conditions and expectations.

Recognizing these industry realities, one can nonetheless examine available evidence. Chart 1 shows AT&T's tariffed monthly rate for a 700 mile T1 (1.54Mbps) circuit. The trend through the 1990s is similar to that for channel terminations: a slight reduction 1990-1995, and a slight increase 1995-1998. Table 6 shows average monthly prices for T-1 (1.54Mbs) and T-3 (44.736Mbs) circuits according to Dataquest. These figures indicate about a 20% reduction 1995-1999. Other data show wider swings. According to a major industry consulting group, from June 1996 to December 1997 MCI's price for a 1000 mile 44.74 Mbps leased line rose 42%.[40]

---

[37] For more discussion of this issue, see Galbi, Douglas, "Transforming Network Interconnection and Transport", forthcoming in CommLaw Conspectus and currently available on the web at http://www.erols.com/dgalbi/telpol/think.htm

[38] More precisely, the tariffs for companies that provide inter-LATA transport services are filed with no justifying documentation and are not closely examined.

[39] Such discounts can be 50% or more. See Ovum, *Future Pricing Trends for Bandwidth*, Final Report (draft) to the National Office for the Information Economy, Australia (Aug 30, 1999) [on the web at http://www.noie.gov.au/projects/information_economy/bandwidth/consultancies.htm ] p. 54 and OECD DSTI/ICCP/TISP(99)4/FINAL, "Building Infrastructure Capacity for Electronic Commerce Leased Line Developments and Pricing" (5 Aug 1999) pp. 19-20.

[40] Rendleman, John, "Connectivity crunch stymies IT access to high-speed lines," *PC Week Online*, 4/13/98, quoting data provided by International Data Corp.

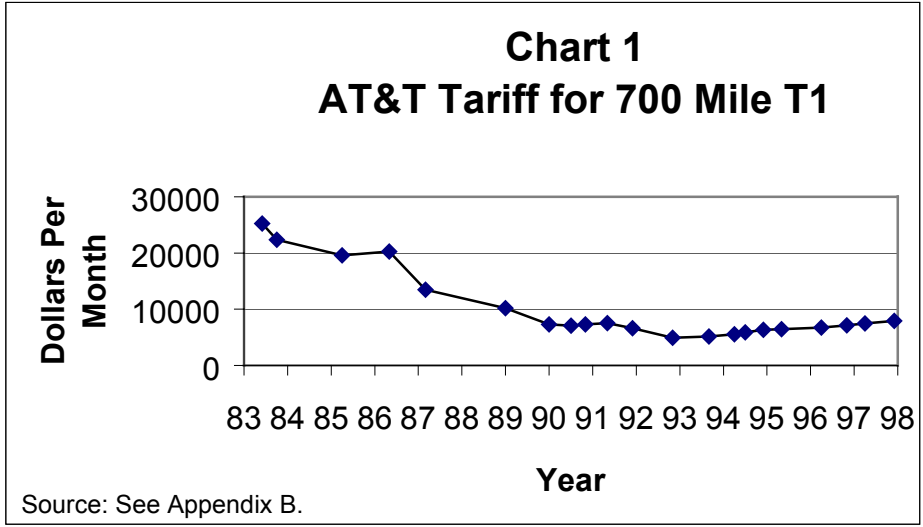

**Chart 1**
**AT&T Tariff for 700 Mile T1**

Source: See Appendix B.

| Table 6 |||
|---|---|---|
| **Long Distance Leased Line Prices** |||
| (dollars per month) |||
| Year | T-1 | T-3 |
| 1994 | 760 | 5,830 |
| 1995 | 760 | 5,830 |
| 1996 | 730 | 5,554 |
| 1997 | 680 | 5,260 |
| 1998 | 650 | 5,000 |
| 1999 | 620 | 4,750 |
| Source: See Appendix B. |||

Price indices from bandwidth exchanges show significant reductions in bandwidth prices for exchange-traded bandwidth. From June 1998 to September 1999 Ratexchange, a bandwidth exchange with relatively strong U.S. domestic operations, shows a 39% decline in its price index for DS3 circuits on major U.S. east-west routes.[41] Chart 2 shows price indices from Band-X, a leading bandwidth exchange with relatively strong European operations.[42] The Band-X world composite price index fell about 63% from October 1998 to July 2000. Larger price reductions occurred on routes with greater bandwidth demand. These price indices are for bandwidth traded on particular exchanges. While bandwidth exchanges are a significant industry development, they currently capture only a small share of bandwidth sales. Thus such price reductions should not be considered representative of aggregate bandwidth price trends.

---

[41] "Continued Bandwidth Rate Leveling", *Phone+ Magazine*, 12/1999. The figures are based on Ratexchange's Price Index.
[42] Data supplied by Band-X. Users who register at http://www.band-x.com can get a description of the method used to construct the index as well as recent index numbers.

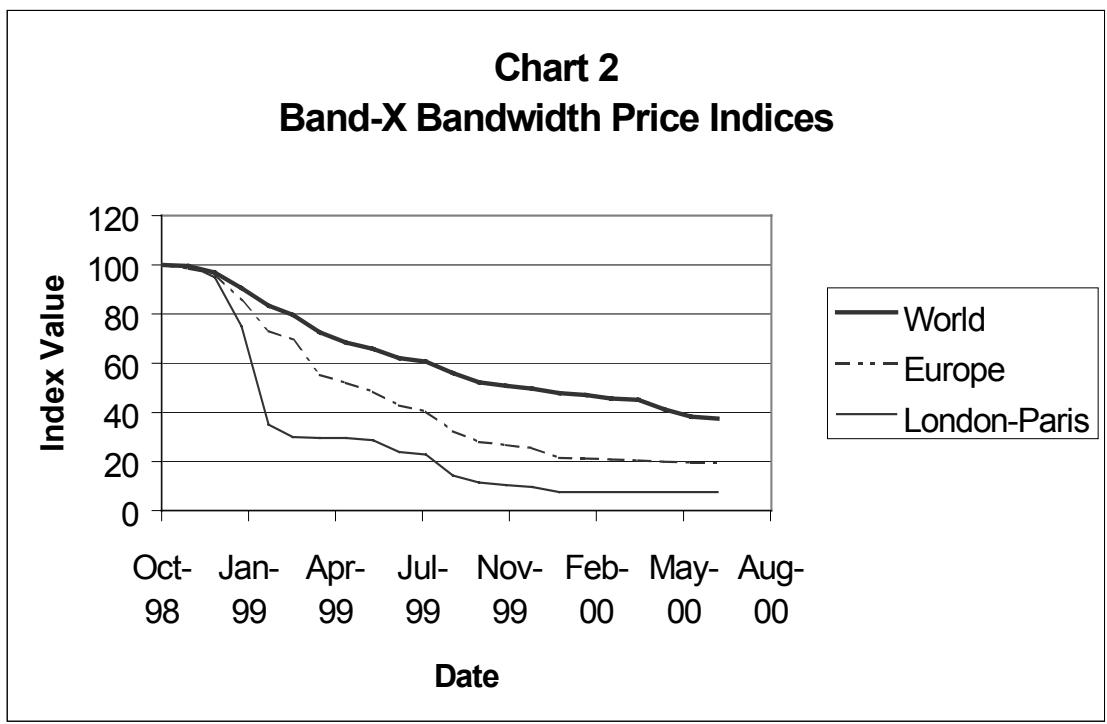

Other more representative evidence from Europe is consistent with relatively constant nominal bandwidth prices. The U.K. regulator has stated that British Telecom's (BT's) prices for 64 Kbps national leased lines have been essentially unchanged in nominal terms since December 1991, while BT's prices for 2 Mbps circuits have fallen only slightly since 1995.[43] Despite the presence of facilities-based competition in the U.K. for over a decade, BT's price for a 300 km 2 Mbps circuit, based on the U.K regulator's estimates for January 1999, was about 40% higher than the price for a similar line in France and Germany and slightly higher than the price for such a line in Portugal.[44] These later three countries have undergone substantial liberalization only since 1998. A survey of U.K. leased line users found that 30% perceived a decrease in leased line prices over the past two to three years, while 50% felt prices had remained the same and 14% perceived an increase in prices.[45] The corresponding figures for the European Union as a whole are 50% perceived a decrease, 33% felt prices had remained the same, and 16% perceived an increase in prices.[46] Despite its ritual repetition, the view that competition has been dramatically driving down bandwidth prices for most users is not a view with much factual support.[47]

---

[43] Oftel, "National Leased Lines, A Further Statement by the Director General of Telecommunications" (Nov. 1999) para. 2.11, 2.15.
[44] Ibid.
[45] Logica Consulting, *Assessment of the Leased Line Market in the European Union*, A Study prepared for the European Commission, Brussels, 19 January 2000 [on the web at http://www.ispo.cec.be/infosoc/telecompolicy/en/Study-en.htm ] p. 70.
[46] Ibid, p. 64.
[47] Concern over the level of leased line prices has recently prompted the EC to issue recommended price ceilings for monthly charges for leased lines. See Commission of the European Communities, Commission

**III. Conclusion**

The growth rate of communications networks and the trajectory of bandwidth prices have not changed significantly in the U.S. through the 1990s. Expectations in the early and mid-1990s of a bandwidth revolution have not yet been fulfilled. The rapid reduction in computer prices in the second half of the 1990s is well connected empirically to the acceleration in U.S. growth and productivity.[48] A rapid increase in bandwidth in use and a rapid reduction in bandwidth prices cannot be connected in the same way to U.S. growth in the second half of the 1990s, simply because such changes did not occur.[49]

When is a rapid, broad-based reduction in bandwidth prices likely to start? One knowledgeable industry observer, while calling past price trends a "great anomaly", predicts that data transmission prices "are likely to start a rapid decline soon."[50] The basis for this optimism seems to be that commodity markets for bandwidth will develop rapidly. Such a development would require a widely recognized structure of coordinating points for network service definition and interconnection.[51] Such a structure appears to be emerging, at least in major cities, despite a policy framework with a much different focus.

While optical technology has been advancing rapidly throughout the 1990s, current economic circumstances make widespread, economically significant use of these technologies more likely. In supplying a wide variety of services, considerable substitution among bandwidth, storage, and CPU cycles are possible.[52] However, such substitution is not perfect, and the vast increases in deployed storage and CPU performance are likely to raise the marginal value of additional bandwidth. In addition, the development of a relatively low transaction-cost market for wide-are bandwidth appears to be necessary to drive network service development and stimulate customer

---

Recommendation, On Leased lines interconnection pricing in a liberalised telecommunications market, Brussels, 24 Nov. 1999, C(1999)3863 [on the web at http://www.ispo.cec.be/infosoc/telecompolicy/en/Main-en.htm ]. In Argentina in 1997, a government decree reduced leased line prices by about 45%, while as of mid-1999 competition has had little effect on leased line prices. See Petrazzini, Ben, and Guerrero, Agustina, "Promoting Internet development: the case of Argentina," *Telecommunications Policy* 24 (2000), pp. 96, 109.

[48] Jorgenson and Stiroh (2000) and Oliner and Sichel (2000) represent the most current work.

[49] The Internet may contribute to growth by fostering common data communications standards that create new business opportunities and encourage productivity-enhancing business re-organization. The aggregate significance of such effects is difficult to measure. Moreover, the price and use of bandwidth is likely to be correlated with the impact of the Internet. In particular, if bandwidth prices drop dramatically and bandwidth use increases dramatically, the effect of Internet technologies is likely to be much larger.

[50] Odlyzko, Andrew, "The history of communications and its implications for the Internet", preliminary version, June 16, 2000, pp. 133, 136 [on the web at http://www.research.att.com/~amo ].

[51] For further discussion of this issue, see Galbi, Douglas, "Transforming the Structure of Network Interconnection and Transport" (2000), forthcoming in *CommLaw Conspectus*, and Galbi, Douglas, "Transforming Network Interconnection and Transport: Policy Direction Summary", forthcoming in *info*. Both papers are currently available on the web at http://www.ssrn.com and http://www.erols.com/dgalbi/telpol/think.htm

[52] Chris Savage has made this point forcefully on numerous occasions on the Cybertelecom discussion list (see http://www.cybertelecom.org/cybert.htm).

demand.[53] Customers are likely to be less interested in bandwidth *per se* than in getting valuable applications that use high bandwidth.[54] The high valuations currently associated with optical networking companies indicate capital markets' belief that necessary changes in industry structure will be forthcoming shortly.

---

[53] One of the surprises of the late 1990s is that users have been relatively slow to upgrade from 28.8Kbps/33.6Kbps modems to 56Kbps modems, which have been widely available since mid-1997. In June 1999 an industry observer noted, "…most people haven't upgraded their paltry 28.8Kbps modems to anything faster despite the fact that 56Kbps modems can be had for less than $100." See Dworak, John C., "Bandwidth Conundrum," Boardwatch, June 1999 [on the web at http://boardwatch.internet.com/mag/00/jun/bwm33.html ]. In January, 2000, 47% of U.S. homes with Internet access were using modems with speeds equal to or less than 33.6Kbps. Another 46% of home users had 56Kbps modems, while the remaining 7% had higher speed access. Based on Nielsen Netratings. See Lake, David, "The Right Tools for the Job," The Standard, March 13, 2000 [on the web at http://www.thestandard.com/research/metrics/display/0,2799,12809,00.html ].
[54] While the trade press is full of reports of high demand for DSL, DSL is an enabling technology that customers are not likely to value in itself.

**Appendix A**
**Further Discussion of Bandwidth Price Trends**

Constructing bandwidth price aggregates from 1990 to 2000 involved considerable effort in data collection and coding. Bell Atlantic (south) and US West were chosen for this analysis because data for them was most readily available in a convenient form.[55] Moreover, Bell Atlantic (south) is a relatively urban east-coast RBOC, while US West has the most rural territory of any RBOC. These two companies account for about one-third of RBOC lines. Industry experience and non-systematic data suggests that aggregated statistics for them are good indicators for the RBOCs as a whole.

Describing price trends for local exchange carriers' leased line offerings involves important standardization, weighting, and aggregation issues. For example, with different volume, term, zone, and distance plans, USWest has in the year 2000 435 different rates for fixed and per mile monthly charges for inter-office DS3 circuits. The number and structure of rates has changed significantly throughout the 1990s. In the figures in Table 4 in the text and Table A2 below, similar types of rates (fixed monthly charge, per mileage charge) are aggregated based on the previous years' demand. Rates have been aggregate across US West and Bell Atlantic (south) also using previous years' demands as weights.

Because the total price for an inter-office circuit depends on its length, and average circuit lengths have changed between years, the prices in Table 4 have been computed based on a fixed, representative circuit length. The circuit lengths used are 13, 15, 10, and 10 miles for VG, DDS 56Kbps, DS1, and DS3, respectively. These lengths are representative figures based on the average circuit lengths for USWest and Bell Atlantic from 1989 to 1999.

Leased line contract duration needs to be considered in constructing a price series. As Table A1 shows, contract durations have changed significantly across the 1990s. Longer duration contracts have lower monthly rates. For DDS, DS1, and DS3 circuits in the year 2000, the discount relative to a monthly contract is 10-15% for a 3-year contract and 20-35% for a 5-year contract. Term discounts for voice grade circuits are much shallower; under 5% and under 10% for 3-year and 5-year terms, respectively. A decrease in contract duration over time would mean that uncorrected prices under-estimate a duration-corrected fall in prices. As Table A1 shows, this is an issue only for DS3 circuits in the second half of the 1990s. Average contract duration for DS3 circuits has fallen from slightly under 4 years in 1995 to slightly under 3 years in 2000. Correcting for this factor in Table 4 would indicate a 10-15% reduction in DS3 prices from 1995 to 2000.

---

[55] Bell Atlantic (south) merged with Nynex, and then with GTE. The resulting company is now called Verizon.

| Year | VG (64Kbps, 128Kbps) | DDS (56Kbps) | DS1 (1.5Mbps) | DS3 (44.7Kbps) |
|---|---|---|---|---|
| 1989 | 1.0 | 1.0 | 4.1 | 46.6 |
| 1990 | 1.0 | 1.0 | 3.8 | 42.6 |
| 1991 | 1.0 | 1.3 | 7.0 | 37.6 |
| 1992 | 1.0 | 2.5 | 4.8 | 37.7 |
| 1993 | 1.4 | 17.4 | 18.3 | 42.4 |
| 1994 | 1.7 | 13.8 | 29.4 | 45.5 |
| 1995 | 1.8 | 16.3 | 27.1 | 42.4 |
| 1996 | 2.1 | 14.3 | 21.8 | 38.0 |
| 1997 | 8.9 | 15.9 | 24.4 | 38.4 |
| 1998 | 10.8 | 17.4 | 28.2 | 37.4 |
| 1999 | 11.5 | 20.8 | 29.4 | 34.1 |

**Table A1
Average Contract Duration
(in months)**

Inter-office circuits are typically purchased from LECs in conjunction with other services such as installation services, maintenance services, multiplexing, cross-connects, and interconnections, and channel terminations. Table A2 gives average prices for channel terminations, which connect an end-user location to a local exchange carrier office. The prices in Table A2 are calculated in the same way as the inter-office figures in Table 4, except there is no per mile charge applied to channel terminations. Average prices for DS3 channel terminations have risen significantly since 1995, while prices for VG and DS1 channel terminations have risen slightly. The overall revenue weighted price fell sharply from 1999 to 2000 because revenue associated with DS1 and DS3 channel terminations grew much faster than that for DDS 56Kbps and VG channel terminations. The rates covered in Table 4 and Table A2 account for about 85% of revenue associated with leased lines. Hence the trends apparent in those tables give a fairly complete picture of the aggregate price trends for LEC leased line services.

## Table A2
## Local Channel Termination Prices
(dollars per month per Mbps)

| year | VG (64, 128Kbps) | DDS (56Kbps) | DS1 (1.5Mbps) | DS3 (44.7Mbps) | Over-all (bw wtd) | Over-all (rev wtd) |
|---|---|---|---|---|---|---|
| 1990 | 300 | 1,622 | 123 | 18 | 177 | 248 |
| 1991 | 295 | 1,592 | 118 | 27 | 137 | 217 |
| 1992 | 305 | 1,707 | 112 | 25 | 107 | 218 |
| 1993 | 294 | 1,637 | 112 | 22 | 89 | 219 |
| 1994 | 299 | 1,366 | 102 | 23 | 78 | 220 |
| 1995 | 319 | 1,601 | 96 | 24 | 76 | 250 |
| 1996 | 319 | 1,444 | 98 | 25 | 77 | 251 |
| 1997 | 320 | 1,471 | 102 | 27 | 80 | 272 |
| 1998 | 311 | 1,453 | 96 | 28 | 76 | 245 |
| 1999 | 323 | 1,488 | 97 | 30 | 77 | 231 |
| 2000 | 320 | 1,383 | 99 | 32 | 74 | 192 |

# Appendix B
# Sources and Notes

## Table 1

Notes: Figure for local incumbents is total for RBOCs, GTE, and Sprint. Other local incumbents probably account for an additional 10% of fiber miles.

Source: Kraushaar, Jonathan M., Fiber Deployment Update, End of Year 1998, Industry Analysis Division, CCB FCC [on the web at http://www.fcc.gov/Bureaus/Common_Carrier/Reports/FCC-state_Link/infra.html ]. The 1999 figure for the RBOCs is from the ARMIS 43-08 data. See http://gullfoss.fcc.gov:8080/cgi-bin/websql/prod/ccb/armis1/forms/armis.hts. The 1999 new local entrants figure is computed from the route miles growth rate from New Paradigm Resources Group, "CLEC Report 2000", as reported in "Appraising the CLEC Landscape", X-Change (June 2000) [on the web at www.x-changemag.com ].

## Table 2

Notes: Figures for 1999 and 2000 Trans-Atlantic and Trans-Pacific bandwidth are estimates. Trans-Atlantic and trans-Pacific bandwidth is based on available capacity. Trade press accounts indicate that trans-oceanic cables constructed in the 1990s have rapidly put in use available capacity. This is not the case with terrestrial capacity. A recent study in Australia found that potential backbone capacity can exceed installed (in use) capacity by 100 to 100,000 times. See Australian Government's *National Bandwidth Inquiry*, Table 5.2, on the web at http://www.noie.gov.au/projects/information_economy/bandwidth/index.htm

Source: For transatlantic bandwidth, see Hsu, Cathy, *1998 Section 43.82 Circuit Status Data* (Dec. 1999) IB FCC, Table 7 [on the web at http://www.fcc.gov/ib/td/pf/csmanual.html ]. RBOC inter-office bandwidth is estimated from Bell Atlantic (south) and US West annual access filings. The given figure represents the average amount of sold LEC inter-office bandwidth in use in the given year. Each link sold separately is included in the aggregate. Direct trunked transport (flat-rated LEC inter-office transport that carriers voice traffic that the LEC switches at its end office) is excluded. Aggregate figure for BA (south) and USWest was scaled to RBOCs using total local private line and special access revenue from FCC ARMIS data. Company figures are broadly in agreement with available bandwidth data from company reports. Anomalous US West figure for 1997 was corrected in light of data in company annual report.

## Table 3

Notes: Year listed is year cable went into operation, with given available capacity.

Source: Hsu, Cathy, 1998 Section 43.82 Circuit Status Data.

## Table 4

See Appendix A.

**Table 5**

Sources: Revenue breakdown by circuit type is based on BA (south) and USWest annual access filings. Interstate tariffed revenue is aggregated from all RBOC annual access filings, while total leased line revenue is the some of local private line and special access revenue from RBOC FCC ARMIS filings.

**Table 6**

Source: Dataquest figures, as cited in Hemenez, Matthew, "A little like Goldilocks", Telephony (June 5, 2000) Tables 1 and 2 [on the web at http://www.internettelephony.com/asp/ItemDisplay.asp?ItemID9506 ].